\documentclass{jsv}
\usepackage{amssyme,equation,bm}
\begin{document}
\draft
\volume{157}
\runtitle{VIBRATIONS IN A CUTTING PROCESS}
\runauthor{J. LIPSKI ET AL.}
\begin{frontmatter}
\title{SURFACE QUALITY OF A
WORK MATERIAL INFLUENCE ON
VIBRATIONS IN A CUTTING PROCESS}
\author{J. L{\sc IPSKI}, G. L{\sc ITAK},
 R. R{\sc USINEK}, K.
S{\sc ZABELSKI}} 
 \author{A. T{\sc ETER}, J.
 W{\sc ARMI\'NSKI} {\sc AND} K. Z{\sc ALESKI}}
\address{
Department of Applied Mechanics, Technical University of Lublin
Nadbystrzycka 36, PL-20-618 Lublin, Poland
}
\received{$23\ October\ 2000,\ and\ in\ final\ form\ 21\ March\ 2001$}
\begin{abstract}
The problem of stability in the machining processes is an important task.
It is strictly connected with 
the final quality of a product. In this paper we consider vibrations of a
tool-workpiece system in a 
straight turning process induced by random disturbances and 
their effect on a product surface. Basing on experimentally obtained
system parameters we have done the 
simulations using one degree of freedom model. The noise has been 
introduced to
the model by the Langevin equation.  
We have also analyzed the product surface shape and its
 dependence on the level of noise.
\end{abstract}
\end{frontmatter}

\section{Introduction}
\noindent
The quality of a final surface in a cutting process is of a natural
interest of industry and technology.
Grabec \cite{1,2} and Gradisek with co-workers \cite{3} analyzed a simple
orthogonal
cutting model and found that,
chaotic conditions of tool-workpiece system,
given by appropriate system parameters, are possible. As they have demonstrated,
the appearance of such chaotic
conditions can have crucial effect on the stability of cutting process.
The chaotic vibrations also were
investigated experimentally by Tansel and others in \cite{4}. 
On the other hand instabilities of cutting process have been known 
for long time as a chatter phenomenon \cite{5,6,7}. The mechanism their appearance  
includes the
 nonlinear self-excitation in the cutting process, which  
leads to vibrations
with larger value of amplitude, beyond the admissible limit. One of the source of
instabilities can be identified in the roughness of a material initial surface, which
 introduces
randomness of
the material resistance during
the dynamic process. Wiercigroch and Cheng \cite{8} have investigated the
influence of noise on the orthogonal cutting system. In
their analysis they started from the spectral
representation of stochastic process. Nevertheless, the most common
treatment of dynamical processes
influenced by noise is the Fokker-Plank approach \cite{9,10,11}. 
Because of some considerable difficulties which it meets by solving in higher dimensions
as well as  for numerical reasons it could  be transformed to the corresponding Langevin equations \cite{10,11}.
Here,
following the 
papers \cite{11,12,13,14}, we use the stochastic Langevin equations with
an additive white noise and  solve the dynamic 
equations of examined system. 
Previous  articles \cite{8,11,13} devoted to cutting process
in
presence of noise, focused rather on the problem of
dynamics and possibilities of bifurcations induced by the random
disturbances. 
Interestingly, Wiercigroch and Cheng [8] and later Przystupa and Litak [11], investigating 
orthogonal cutting process with two degrees of freedom,  claimed that in some conditions weak noise can even
stabilize the chaotic attractor.
On the other hand paper \cite{14} deals with the reconstruction of the system dynamics
from the stochastic time series. In their treatment the cutting process was assumed
to be deterministic  but  the measured data were influenced by noise
coming from the measurement procedure.
 
Our paper is also  a contribution
to a complicated problem of dynamics 
of a cutting process, but we focus on the final quality of 
a product surface, and in this context, in the
stability of process. Note, in Fig. 1, the shape of an initial
surface of a cut workpiece. The shape has numerous imperfections which
we will model by  random deviations from the ideal cylindrical surface.  
Adopting a
simple one degree-of-freedom model of regenerative cutting \cite{15}
we included the effect of a previous
pass of a straight turning process
 by a time delay term \cite{16}.

\section{Deterministic Model of Cutting process}

The physical model of a straight turning process, corresponding to our experimental system, is presented
in Fig. 2. Here we have introduced
the following notations: $v_f$ is a relative velocity between the tool and the workpiece, $h_0$ is an
assumed
initial while $h$ an actual cutting
depth; $w$ is a principal axis of relative vibrations; 
$y$ indicates the direction normal to the axis of a workpiece
symmetry, $\kappa$ is a tool cutting edge angle;
$k$ and $c$ are the stiffness and damping of the system respectively; $n$ denotes
a rotational velocity of a workpiece;
$f$ denotes the direction of feed in a straight
turning; m is the effective mass of the system.

The main vibration in $w$ direction, perpendicular to the cutting edge, (Fig. 2) and to be precise we should
analyze vibrations as well as  cutting force in $w$ direction.
However  we are interested in the final surface profile given by the time history  of $y$
and not by the actual profile $w$.  
To analyze vibration in $y$
we have done  simultaneous 
projecting vibrations and  forces into $y$ direction.
Thus the deterministic equation of motion of dynamical system, projected on a
normal (to final surface) direction $y$, can be written as
follows \cite{15}:
\begin{equation}
\ddot y + 2\tilde n \dot y + p^2 y = \frac{K}{m} g_y(h,v_f),
\end{equation}
where $p$ is  a natural frequency of free vibrations of the workpiece $p^2=k/m$ while $\tilde n=
c/m$.
 Nonlinearities, appearing in that system, 
are included in the $g_y$ function
\cite{1,2,3,8,15}:
\begin{equation}
g_y(h,v_f)= \left[ c_2 \left( \left| \frac{v_f}{v_0} \right| - 1 \right)^2
+1 \right] \left[ c_3 \left( \frac{h}{h_0} -1 \right)^2 +1 \right]
\frac{h}{h_0} \Theta (h) {\rm Sgn} (v_f). 
\end{equation}
Cutting depth $h$ and relative velocity $v_f$  are defined \cite{14,15}:
\begin{equation}
h=h_0+ y(t') -y(t),~~~~v_f=1-\frac{\dot y}{v_0}.
\end{equation} 
$\Theta(h)$ and ${\rm Sgn} (v_f)$ correspond to step functions: Heaviside
and sign
functions respectively and $v_0$ ia the  linear velocity of a rotational 
motion of a workpiece during  a steady cutting process. 

$t'$ is the time of
a previous pass:
\begin{equation}
t'=t-\Delta t,
\end{equation}
where $\Delta t$ is a workpiece revolution time during machining.
The shape of a nonlinear function $g_y$ (Eq. 2) dependent
on $h$ and $v_f$  is presented in Fig. 3. Note the two-dimensional
surface $g_y=g_y(h,v_f)$
was plotted only for positively defined $h$.
In case of negative $h$, the force on the left hand side of Eq. 1, is zero
because of the contact loss between
the tool and a workpiece. The sudden sign change
 of a cutting force in a function of  relative velocity $v_f$ 
is due to a  friction
phenomenon between the tool and a chip.

Our model (Eqs. 1-3) with one degree-of-freedom is a serious simplification of
a physical situation.
However our aim is not the comprehensive description of a cutting process.
Here we want to concentrate on particular aspects of it. In spite of
simplicity of the examined model   still allows the chatter vibrations to be generated due to
the nonlinearities in of the cutting force $g_y(h,v_f)$ as it was shown in paper \cite{15}.
In our model chatter is  generated  by
a combination of the friction phenomenon between the tool and chips, and
the impact of a tool after  loosing its contact with a workpiece.
Warmi\'nski and others \cite{15} examined the second pass of the orthogonal cutting process 
by using similar     
model. The results
obtained there have indicated that such model can lead to periodic, quasi-periodic as well as chaotic
vibrations due to the initial harmonic modulation of the machined surface.   

For a numerical calculation purpose we have written the equations (Eqs. 
1-4) in
discrete way introducing the constant
time step $\tau$:
\begin{eqnarray}
t_{r+1} &=& t_r + \tau, \nonumber \\
y_{r+1} &=& y_r +v_r \tau, \\
v_{r+1} &= &v_r + (-2 \tilde n v_r - p^2 y_r + \frac{K}{m} g_{yr}) \tau,
\nonumber
\end{eqnarray}
where $t_r$ is a sampling discrete time
after $r$ time steps.
The function $g_{yr}$ should be expressed:
\begin{eqnarray}
g_{yr} &=& g_y (h_r, v_{fr}), \nonumber \\
h_r &=& h_0+y_s -y_r, \\
v_{fr} &=& 1- \frac{v_r}{v_0}, \nonumber 
\end{eqnarray}
where $r$ and $s$ are  natural numbers. 
The time difference between $y_r$ and $y_s$  coordinates;  $\Delta t =(r-s)
\tau $
relates to the time  of workpiece revolution (Eq. 4). The system
parameters
obtained from the experiment are
following: $p=785$ rad/s, $m=12.1$kg, $K=620$ N, $h_0=1.5 \times 10^{-3}$
m,
$f=0.1~\times 10^{-3}$ m/rev, $2 \tilde n =190~ $1/s, $v_0=0.1$ m/s $\kappa=
70^o$ and $c_2=0.5$,
$c_3=1.55$ are cutting process constants derived from \cite{7,15}.

\section{INFLUENCE OF NOISE}

Most of real dynamical processes are disturbed by the random signal. In
case
of cutting process they come through the
roughness of the initial surface (Fig. 1). Other sources of such
disturbances can be found in  a spontaneous breaking of chips and
the couplings of the tool and the workpiece to other
dynamic parts of the experimental standing.

To describe the stochastic system we introduce to the model a random
component by an additive white noise of Gaussian distribution \cite{10,11,12,13,14}.
Usually stochastic dynamic systems are investigated by using the Fokker-Planck equation
\cite{8,9,10,11}.
One dimension version of it reads:
\begin{equation}
\frac{\partial}{\partial t} P(y,t) = \frac{\partial}{\partial y} \left[ \nu (y)
P(y,t)\right] + D \frac{\partial^2}{\partial y^2} P(y,t),
\end{equation} 
where $D$ denotes the diffusion coefficient,  $\nu (x)$ is, in general,  the non-linear drift term
(driving
force) and $P( x,t)$ the probability distribution function. 
As  solving Fokker-Plank equation meets some considerable difficulties \cite{10,11}
we have transformed it into the corresponding Langevin equation: 
\begin{equation}
\dot y = z(y) + g \Gamma (t),
\end{equation}
where  $g \Gamma (t)$ is a Gaussian distributed random 'force' with the strength $g$ and 
$\Gamma (t)$ is assumed to satisfy:
\begin{eqnarray}
< \Gamma(t) >  & = & 0, \\
< \Gamma(t), \Gamma(t') > & = & 2 \delta( t-t'), \nonumber
\end{eqnarray}
where the brackets denote an average over the probability distribution function.
Starting with the definitions of the drift $\nu$ and diffusion $D$ coefficients obtained by
the Kramers-Moyal expansions in the derivation of Fokker-Plank equation  from a
Chapman-Kolmogorov equation \cite{17}, we can find the relation between $\nu(y)$ and $D$ of the
  Fokker-Plank equation (Eq. 7) with $z(y)$ and $g$ of the Langevin equation (Eq. 8). Thus
the
Langevin equation can be finally expressed by drift and diffusion terms of the
initial Fokker-Plank 
equation as:
\begin{equation}
\dot y = \nu(y) + \sqrt{D} \hat \Gamma (t)
\end{equation}
For the actual numerical calculation we have discretized form:
\begin{equation}
y(t+\tau)-y(t)= \int_{t}^{t+\tau} {\rm d} t' v(t') + \sqrt{D}
\int_{t}^{t+\tau} {\rm d} t' \Gamma (t') \approx \nu (t) \tau +
\sqrt{D} \hat  \Gamma (t), 
\end{equation}
where
\begin{equation}
 \hat  \Gamma (t) =\int_{t}^{t+\tau} {\rm d} t' \Gamma (t')
\end{equation}
is a superposition of Gaussian distributed random numbers which again are
of a Gaussian form. Namely:
\begin{equation}
\hat  \Gamma (t) = a \omega (t).
\end{equation}
In our case the average value of $\omega$  has been chosen as $<\omega>
=0$ while
its variance as
$<\omega^2>=2$, respectively. From  the integration of the Gaussian (Eq. 12) the
coefficient
$a$ (Eq. 13) depends on the time integration step
$\tau$ via  $a  = \sqrt{\tau}$.
Equation 8 can be, in general, 
solved by higher order
algorithms like Runge-Kutta one \cite{18}. However here, for simplicity,
we
have
limited our discussion to the simplest algorithm of Euler type. Thus, the final
form of
the Langevin equation in the lowest order perturbation, suitable for
numerical
integration is given by the following
expression:
\begin{equation}
y_{r+1}=y_r + \nu (y_r)\tau + \sqrt{D} \sqrt{\tau} \omega (t_r).
\end{equation}                                                               

Here we analyzed the one-dimensional version of the Langevin and Fokker-Plank equations.
The similar discussion on the $m$-dimensional stochastic Langevin equation and its relation
to 
the corresponding Fokker-Plank equation can be
found in 
\cite{14}.

In our model, Eq. 9 for $y_r$ has to be supplemented by the rest of
equations  (Eqs. 5,6) for discrete time $t_r$ and velocity $v_r$ 
which is now substituted by the corresponding drift term $\nu_r$: 

\begin{equation}
\nu_{r+1} = \nu_r + (-2 \tilde n v_r - p^2 y_r + \frac{K}{m} g_{yr}) \tau.
\end{equation}

Using the above procedure (Eqs. 4,5 and Eq. 14-15) we have done the
simulations
for
a constant time step $\tau =0.741 \times 10^{-4}$ s,
corresponding to the workpiece revolution time $\Delta t=0.741 \times 10^{-1}$ s
 and a number of diffusion constants values $D$. Figures 4 a--d show the
time
histories of $y$ of initial 3 seconds
of cutting work for experimentally identified system parameters (Sec. 2).
For a deterministic system ($D=0$, Fig. 4a)
we observe the stable cutting
 process with no vibrations. Figures 4b, c relate to cutting process in
presence of noise. The diffusion constant values
in these figures are $D=10^{-5}$ and $D=10^{-4}$ respectively. One can see
easily the presence
of small vibrations (Fig. 4b), which
are growing with increasing of the noise level (Fig. 4c). Obviously, such vibrations have a
significant effect on the quality of
a workpiece surface. It is shown in
Fig. 5, where the error shape of a  surface is plotted as a function of
a workpiece rotation angle after 3 s of cutting
work.

Modulation of shape caused by random disturbances depends on  a noise
level.
Both:
input and output random signals can be
easily measured by standard deviations. For various $D=10^{-6}$, $10^{-5}$
and $10^{-4}$
we have got the following values of
standard deviations: $\sigma= 3.79 \times 10^{-
7} $ m, $1.21 \times 10^{-6}$ m and $4.16 \times 10^{-6}$ m. In
Figs. 6a and b we have compared
the input and output random signals for
one of the above cases ($D=10^{-5}$). Figure 6a shows the distribution of
Gaussian disturbances of input noise $\omega$ (Eqs.
8,9) while Fig. 6b corresponds to the errors
 of workpiece shape after cutting. In Fig. 6b deviation
from the normal probability
distribution is caused by nonlinear dynamics of the cutting process. To
quantify the system answer we have used
standard deviations of the output signal
$\sigma$ as a function of diffusion constant $D$. It is plotted in
the logarithmic
scale in Fig. 7. We have checked that   
$\sigma (D)$ can be scaled as a square root as far as noise level is low
while
for stronger noise its effect on the
fluctuations of $y$ is stronger.

\section{SUMMARY AND CONCLUSIONS}
We have  considered  vibrations of a tool-workpiece  system in a straight turning
process induced  by
random disturbances and their effect on a product surface. 
Using a single degree-of-freedom model we have focused on the 
combined effects of the friction nonlinearities and tool-workpiece contact loss. 
We have
noticed that for large
enough level of noise, the tool and a workpiece start to
vibrate due to random forcing. Such excitation can interact with a complex
dynamics of the system leading
to process non-stability and,
in the end, to much  worse final quality of the machined product.
In case of  a relatively small level of noise (a small value of the diffusion constant $D$) the surface
shape
error scales as a square root of $D$. For a higher value of a noise level the shape error is proportional to $D$.
Clearly,  in a straight turning the initial surface roughness influence the quality of a final product.
This is the principal result of our paper, which lead to a conclusion that 
one has to  prepare workpiece which initial surface satisfies the appropriate criteria.

\begin{ack}
The work has been partially supported by Polish State Committee for
Scientific
Research (KBN) 
under the grant No. 126/E-361/SPUB/COST/T-7/DZ 42/99.
We would like to thank the organizers of $3^{rd}$ International Symposium
"Investigation of Nonlinear Dynamic Effects in Production Systems" 26-27
September 2000 in Cottbus (Germany)
for giving two of us (J.L. and R.R.)  opportunity to present this work.
We would like to thank the unknown referee for  
 valuable comments  and Dr W. Przystupa for helpful discussions.
\end{ack}

\newpage

{\center FIGURE CAPTIONS}

\noindent
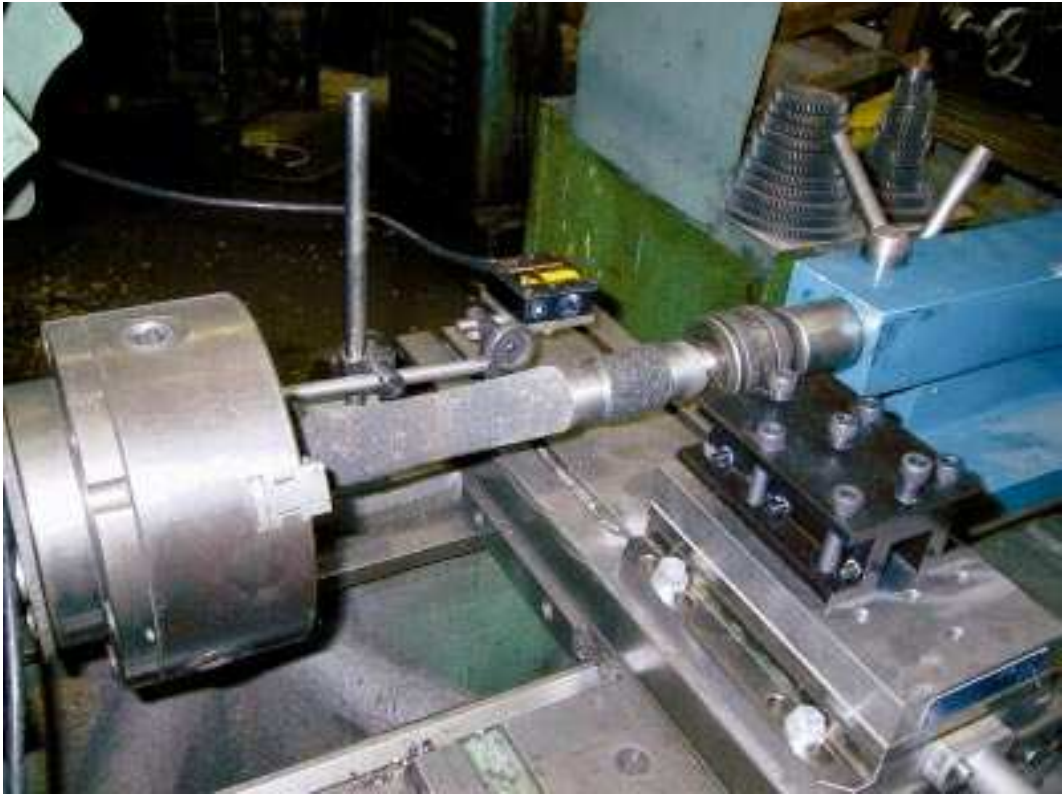
Figure 1. Experimental standing with a cast iron workpiece. \\
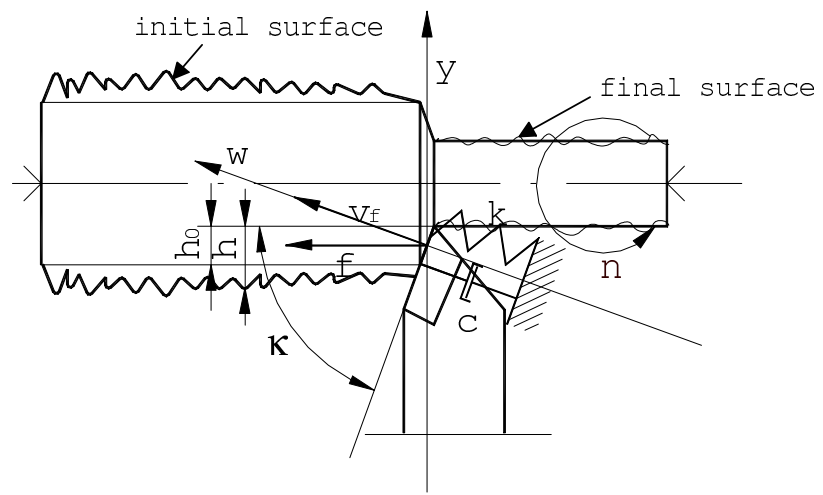
Figure 2. Physical model of a straight turning process. \\
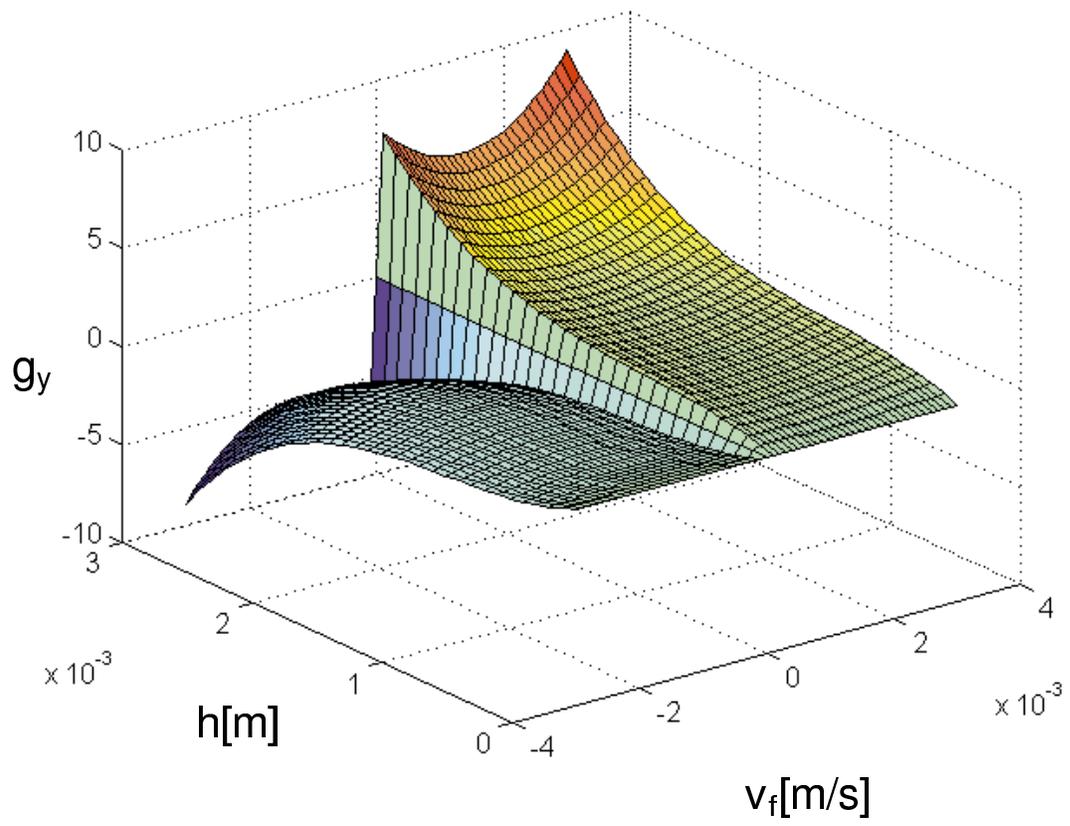
Figure 3. Nonlinear function $g_y(h,v_f)$ versus cutting relative
velocity. \\
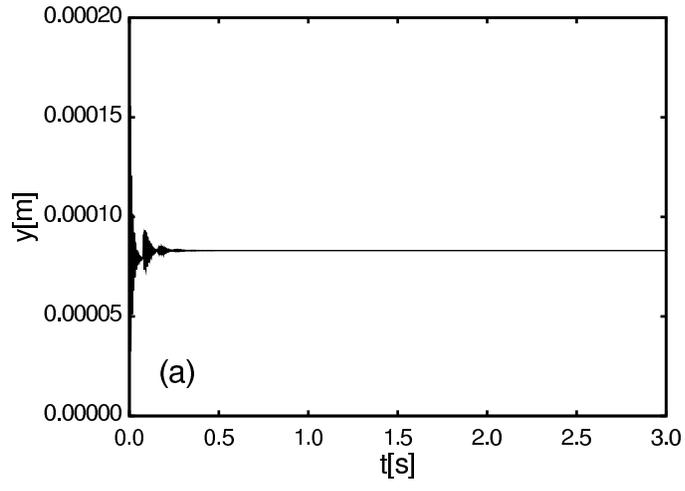
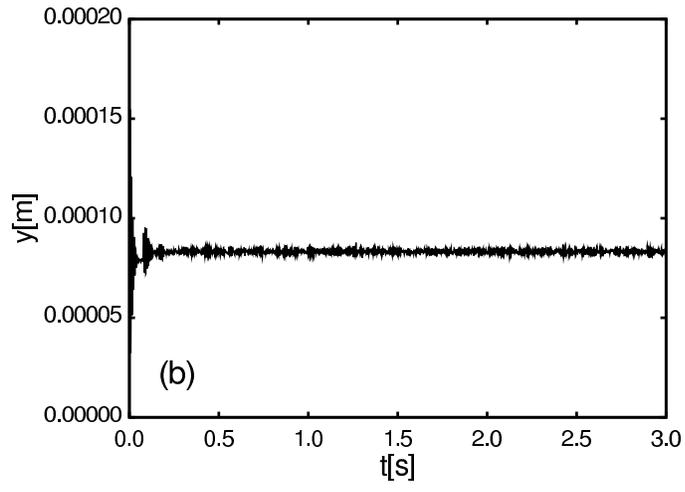
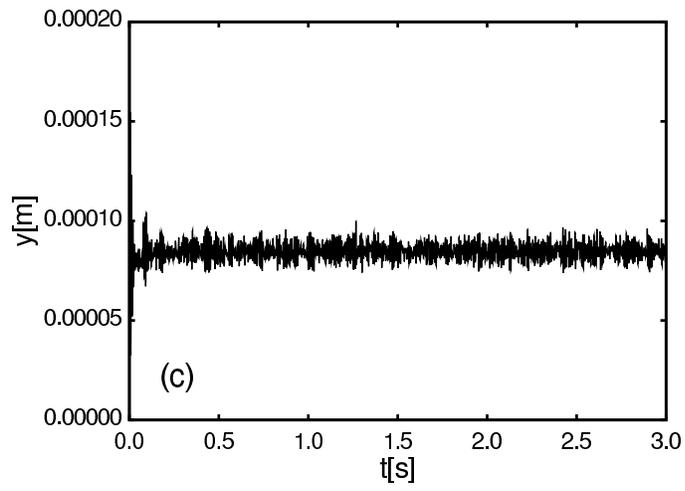
Figure 4. Time histories of $y$ for various values of diffusion constants
(a) $D=0$, (b)  $D=10^{-5}$, (c)  $D=10^{-4}$. \\
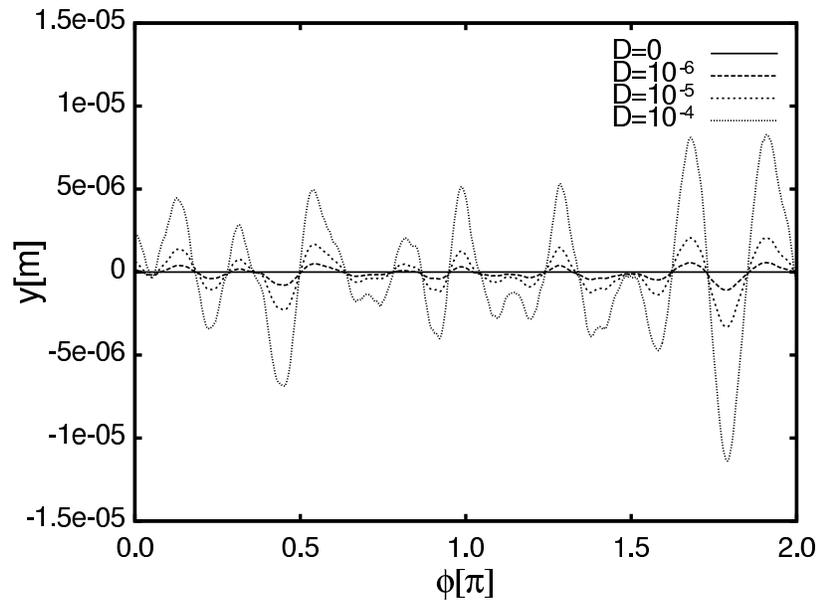
Figure 5. Shape error as a function of workpiece rotation angle after $3
s$ of cutting. \\
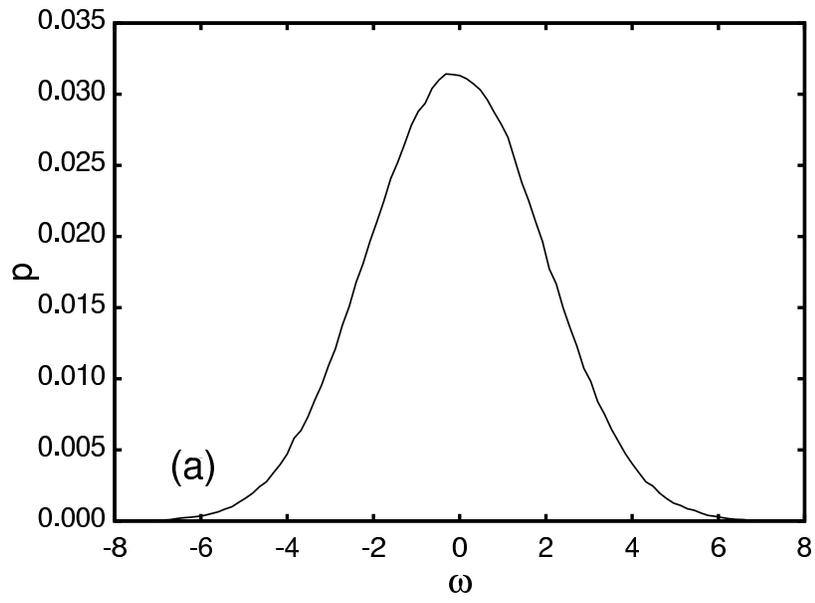
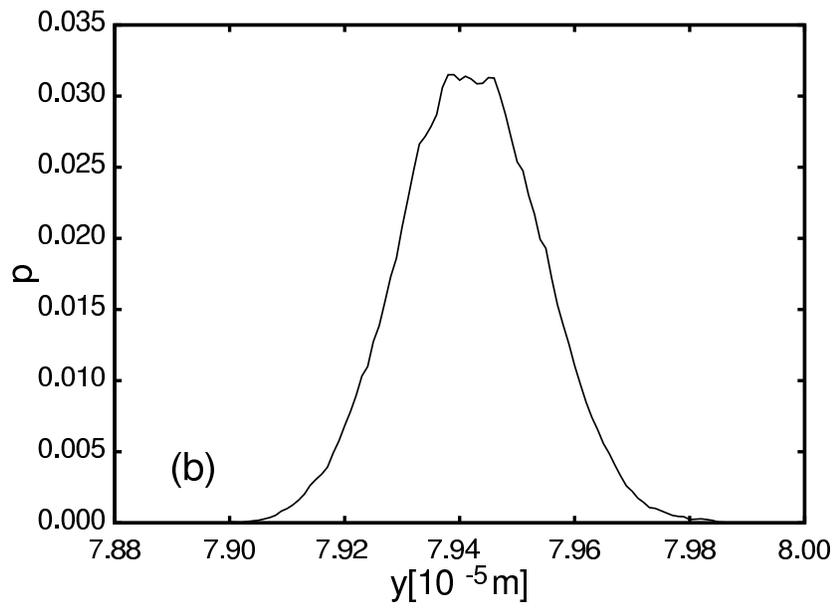
Figure 6.  Distribution probabilities of random input (a) and output $y$
(b) signals of the model for $D=10^{-5}$. \\
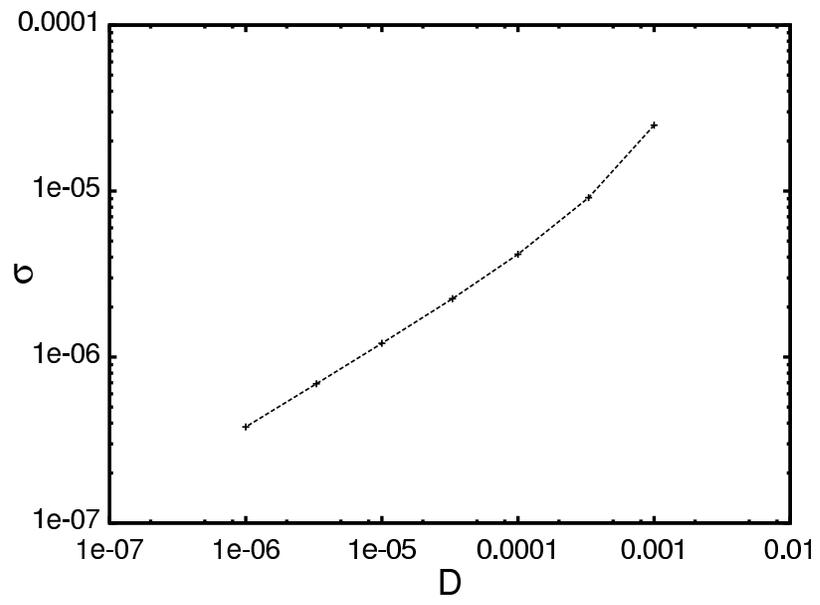
Figure 7. Standard deviation $\sigma$ as a function of a diffusion
constant
$D$.

\end{document}